\documentclass[10pt]{article} 
\usepackage{setspace}
\usepackage[a4paper, top=2cm, bottom=2cm, left=2cm, right=2cm, marginparwidth=2cm]{geometry}
\usepackage{amsmath} 
\usepackage{units} 
\usepackage{placeins} 
\usepackage{graphicx}

\begin{document}

\begin{center}

\textbf{\fontsize{22}{\baselineskip}\selectfont Fast, multicolour optical sectioning over extended fields of view by combining interferometric SIM with machine learning}
\bigskip

Edward N. Ward,$^1$ Rebecca M. McClelland,$^1$ Jacob R. Lamb,$^1$ Roger Rubio-Sánchez,$^{1,2}$ Charles N. Christensen,$^1$ Bismoy Mazumder,$^1$ Sofia Kapsiani,$^1$ Luca Mascheroni,$^1$ Lorenzo Di Michele,$^{1,2}$ Gabriele S. Kaminski Schierle,$^1$ and Clemens F. Kaminski$^1$

\bigskip

$^{1}$Department of Chemical Engineering and Biotechnology, University of Cambridge, Cambridge, CB3 0AS, UK\\
$^{2}$fabriCELL, Molecular Sciences Research Hub, Imperial College London, London, W12 0BZ, UK\\

\end{center}
\section{Abstract}

Structured illumination can reject out-of-focus signal from a sample, enabling high-speed and high-contrast imaging over large areas with widefield detection optics. Currently, this optical-sectioning technique is limited by image reconstruction artefacts and the need for sequential imaging of multiple colour channels. We combine multicolour interferometric pattern generation with machine-learning processing, permitting high-contrast, real-time reconstruction of image data. The method is insensitive to background noise and unevenly phase-stepped illumination patterns. We validate the method \textit{in silico} and demonstrate its application on diverse specimens, ranging from fixed and live biological cells to synthetic biosystems, imaging at up to \unit[37]{Hz} across a 44~$\times$~\unit[44]{$\mu m^2$} field of view. 

\section{Introduction}

Widefield fluorescence microscopy permits the imaging of biological structures with a high specificity, however, out-of-focus light limits image contrast. Unless planar excitation profiles are used for illumination – such as in lightsheet \cite{Huisken2004OpticalMicroscopy} or HiLo \cite{Tokunaga2008HighlyCells} microscopy – fluorescent probes located above and below the focal plane contribute to the signal collected in the final image. 

Scanning confocal microscopy circumvents this problem through the use of a pinhole to physically reject out-of-focus signal. While this is effective at increasing contrast, only a single point in the sample can be imaged at a time and images must be built up sequentially, pixel-by-pixel, greatly increasing acquisition time. Furthermore, high excitation powers are required to compensate for the signal lost through the pinhole. 

A different approach to achieve optical sectioning (OS) is through structured illumination microscopy (SIM) \cite{Neil1997MethodMicroscope}. Here, a fluorescent sample is illuminated by patterned excitation light and the emitted fluorescence is imaged with widefield detection. In super-resolution SIM, interference patterns are produced by the interaction of the patterned excitation with the structures of the sample and these interference patterns are used to extract high-resolution information about the sample \cite{Heintzmann1999LaterallyGrating, Gustafsson2000SurpassingMicroscopy}. OS-SIM makes use of the fact that the modulation depth of the excitation pattern is highest in the in-focus plane and decreases rapidly with distance from it. Hence, the in- and out-of-focus structures can be distinguished by differences in stripe contrast. To achieve this, the sample is typically illuminated with a sinusoidal stripe pattern, and three sequential images are acquired as the pattern is shifted in phase over the sample. Under these conditions, only the in-focus structures will show a change in intensity between the phase shifted images and therefore these can be extracted and the three raw images reconstructed into one optically sectioned image. This method was pioneered by Neil \textit{et al.} \cite{Neil1997MethodMicroscope} who used the squared difference (SD) between the three phases to reconstruct the image: 

\begin{equation}
I_{R} ^{2} = (I_1 - I_2)^2 + (I_2 - I_3)^2 + (I_1 - I_3)^2,
\label{eq:refname1}
\end{equation}

where $I_R$ is the reconstructed image and $I_{n}$ represents the n-th phase image of the sample under striped illumination with the phase shifted by $2(n-1) \pi $/3. Phase stepping in this way ensures a uniform average illumination and prevents striping artefacts in the reconstructed image. While simple to implement, this method is highly sensitive to noise, which becomes amplified in the SD reconstruction. Additionally, even small movements of the sample or deviations from the ideal phase step size introduce reconstruction artefacts. While recent efforts have improved on this early method, reconstructions remain prone to artefacts which require substantial filtering and post-processing \cite{Li2020FastVivo}. 

To address these issues, we propose a method for OS-SIM which combines interferometric pattern generation with machine learning (ML): ML-OS-SIM. We use two complimentary neural networks to reconstruct OS-SIM data: a fast network for real-time reconstructions and a heavier network for post-acquisition reconstructions. In combination with an interferometric SIM setup \cite{Ward2022MachineImaging}, we achieve OS-SIM reconstructions in multiple colour channels simultaneously. Using the fast-reconstruction network, we demonstrate on-the-fly processing of OS-SIM data during acquisition, allowing users to immediately and intuitively visualise reconstructions as they navigate the sample. The software to achieve this is packaged in a user-friendly GUI which, combined with the simplicity of the interferometric SIM design, makes the method easy to implement on existing microscope setups.

\section{Methods}     

\subsection{Hardware implementation}
For optimal reconstructions of OS-SIM data, the spatial frequency of the striped illumination pattern must be half of the maximum  spatial frequency observable by the microscope \cite{Neil1997MethodMicroscope}. As the resolution limit changes with wavelength, the periodicity of the pattern must also be changed when switching between colours. In SIM setups based on spatial light modulators, digital micro-mirror devices or diffraction gratings, this means that only a single colour can be imaged at a time, slowing down image acquisition and, crucially, introducing a temporal offset between acquisitions of different colour channels. To overcome these issues, we use an interferometric method to generate the fringe patterns (Fig. S1) \cite{Ward2022MachineImaging}. This method has the advantage that the periodicity of the structured illumination fringes depends on the wavelength of the excitation light, meaning the fringe spacing is optimised for all wavelengths simultaneously. The setup makes use of a Michelson interferometer and phase stepping is achieved by laterally sweeping the pattern across the sample with a galvanometric mirror element. In our setup, fluorescence signal from the sample is collected and split into three colours using an image splitting device, permitting the simultaneous imaging of multiple colour channels. A \textit{Python} graphical user interface enables hardware control and display of reconstructed images in real time (Fig. S3). All source code is available in a GitHub repository \cite{Ward2023ML-OS-SIM}. 

\subsection{Machine learning networks}

Machine learning (ML) has become a powerful enabler for the pre-processing, post-processing and reconstruction of super-resolution SIM data \cite{Chen2023SuperresolutionReview}.
Compared to classical methods, ML can increase reconstruction quality and does not require the estimation of system parameters. ML has also been applied to the reconstruction of OS-SIM data \cite{Zhang2018DeepMethod, Chai2021DeepMeasurement}, but the impact of sample motion and uneven phase steps on reconstruction quality has not been addressed. Here, we leverage recent advances in ML and optimise two distinct neural networks specifically for the reconstruction of interferometrically generated OS-SIM data. By using two networks, we are able to achieve real-time reconstructions at the point of acquisition, and high-fidelity reconstructions post-acquisition, yielding optimised results even for moving samples.

In the first instance, ML-OS-SIM reconstructions were performed using a lightweight convolutional neural network (CNN) based on the residual channel attention network (RCAN) architecture (Supplementary note 4B) \cite{Zhang2018ImageNetworks}. This RCAN model has a low memory footprint and can perform reconstructions on low-end consumer graphics cards at rates compatible with live reconstructions during imaging.

For high-fidelity image reconstruction post-acquisition, we use video super-resolution (VSR) reconstruction, a video transformer network based on the shifted-window architecture (Supplementary note 4A) \cite{SwinIR2021Liang, Dosovitskiy2020AnScale}. While more computationally intensive, this network is specifically optimised to reconstruct data from moving structures, mimicking dynamic biological samples \cite{Christensen2022Spatio-temporalMicroscopy}. 

To mitigate the challenges posed by training on experimentally acquired OS-SIM data, we simulated large datasets and employed transfer learning to train the network. This approach offers several advantages: firstly, for supervised learning, the ground truth is known, ensuring that the networks do not learn to generate the artefacts produced by classical reconstruction algorithms. Secondly, large and diverse training datasets can be generated, enabling the models to learn the reconstruction process without overfitting to limited experimental datasets. Thirdly, it is possible to train the network to be robust to the specific challenges associated with reconstructing interferometric OS-SIM data (Supplementary note 4C). This is particularly important when the phase changes of the pattern differ from the ideal, evenly spaced phases required for other reconstruction methods to work.

\section{Results and Discussion}

\subsection{Machine learning reconstructions suppress noise and out-of-focus light from volumetric data}

\begin{figure}[htbp]
\centering
\includegraphics[width=0.99\textwidth]{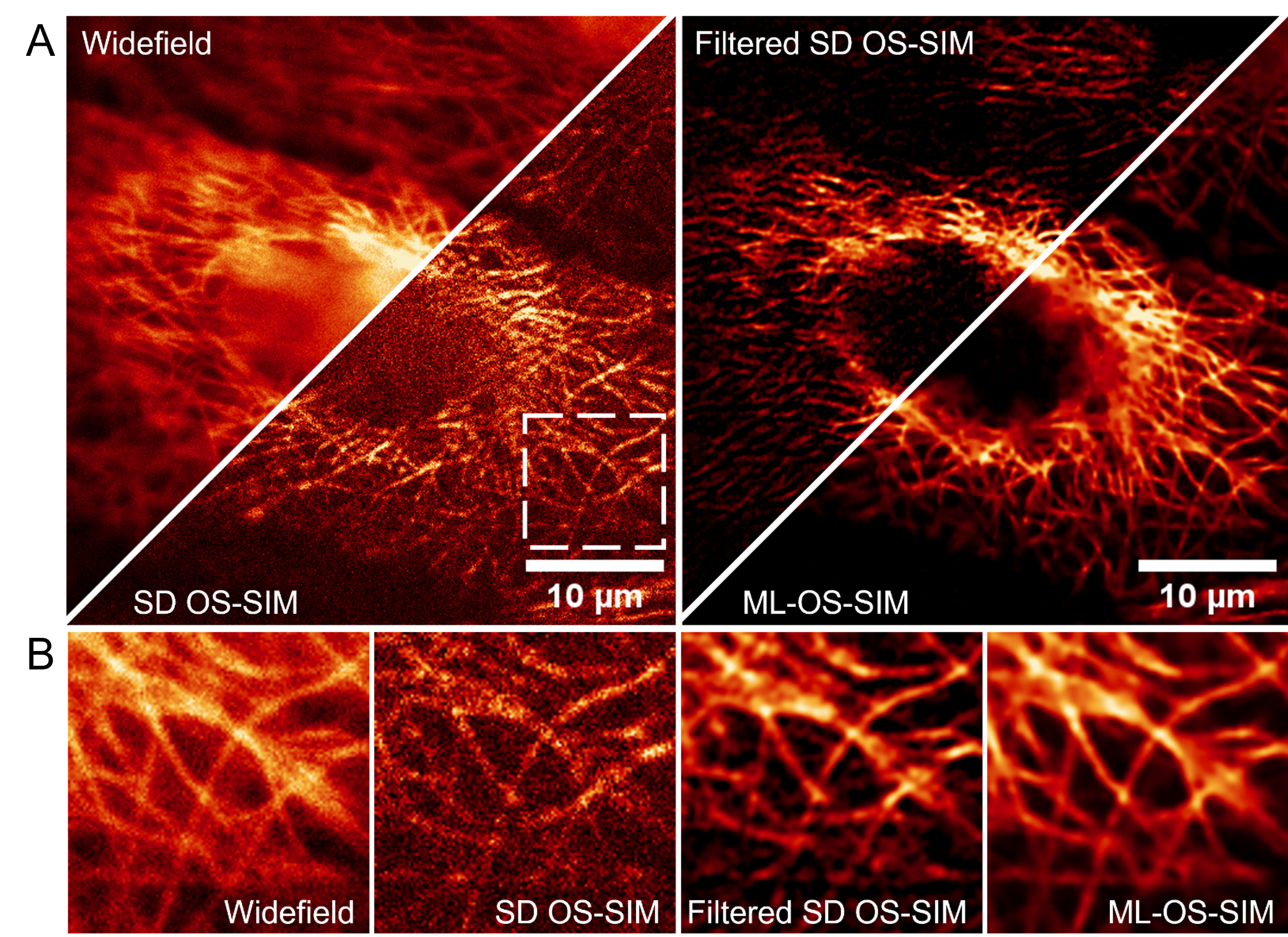}

\caption{Machine learning outperforms classical methods for the reconstruction of optical sectioning structured illumination microscopy (OS-SIM) data. A: Widefield image and reconstructions of OS-SIM data acquired of immunostained $\beta$-tubulin in fixed Vero cells. Square difference (SD) \cite{Neil1997MethodMicroscope} and filtered SD \cite{Li2020FastVivo} both show an improvement in contrast and background rejection, however, both techniques are heavily impacted by noise. In the SD reconstruction, the nose is amplified and the details of the sample are obscured. The filtered SD reconstruction removes this noise but at the expense of artefacts being introduced into the reconstruction. In comparison, ML-OS-SIM reconstruction successfully removes both the background and noise. Scale bars = \unit[10]{$\mu m$}. B: Magnified 10~$\times$~\unit[10]{$\mu m^2$} region indicated on A.} 

\label{fig:reconstruction comparison} 
\end{figure}

We first validated the ML reconstruction technique on volumetric data from fixed biological structures (Fig. \ref{fig:reconstruction comparison}). OS-SIM data from a fixed sample of Vero cells were reconstructed into optically sectioned images using ML and compared to the SD and filtered SD approaches \cite{Neil1997MethodMicroscope, Li2020FastVivo}. All OS-SIM methods exhibit a reduction in the out-of-focus signal, however, the ML approach performs much better in the presence of background noise. To quantify this improvement, the ML reconstructions were tested on simulated biological samples (Figs. S3 and S4). Successful reconstructions of OS-SIM data were possible at approximately one fifth of the signal level where SD approaches fail.

\subsection{ML-OS-SIM can image faster than point scanning confocal microscopy with similar imaging performance} 

\begin{figure} [htbp]
\centering
\includegraphics[width=0.99\textwidth]{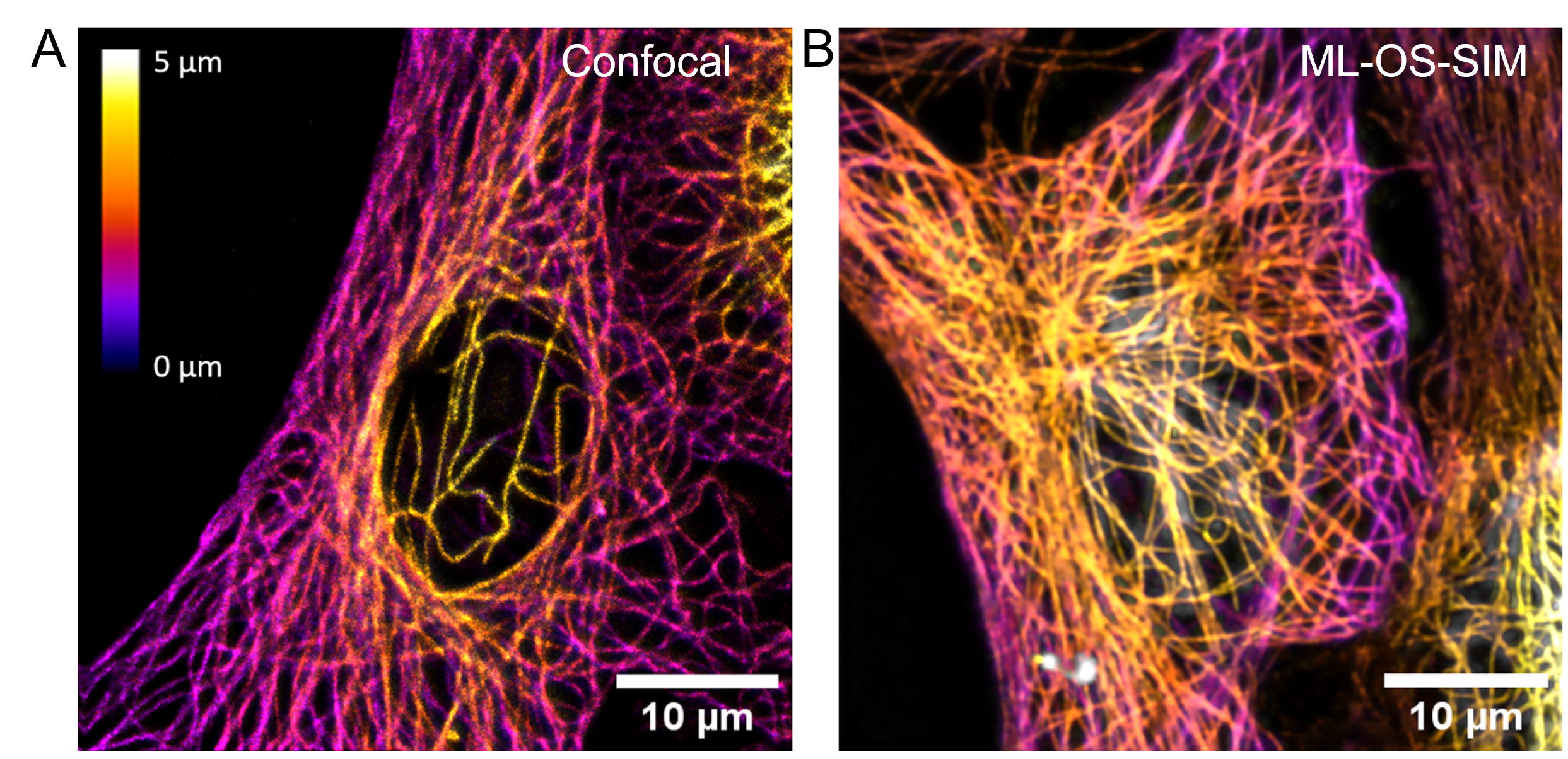}          
\caption{Machine-learning optical sectioning structured illumination microscopy (ML-OS-SIM) produces optically sectioned volumes equivalent to scanning confocal microscopy at faster imaging speeds. Images show immunostained $\beta$-tubulin in fixed Vero cells imaged with A: deconvolved scanning confocal microscopy and B: ML-OS-SIM. Both techniques enable the 3D structure of the microtubule cytoskeleton to be resolved. Using ML-OS-SIM, the complete equivalent volume could be imaged in < \unit[2]{s} compared to \unit[2]{min} \unit[36]{s} for scanning confocal microscopy. Colourmap indicates the z position. Scale bars = \unit[10]{$\mu m$}.}
\label{fig:ConfocalvsOSSIM} 

\end{figure}

To demonstrate the speed improvement possible with interferometric ML-OS-SIM, the system was compared to point scanning confocal microscopy (Fig. \ref{fig:ConfocalvsOSSIM}). The camera exposure time was adjusted to produce reconstructions of a similar quality to confocal images of the same sample. In both images, background light is effectively removed and the 3D microtubule network of the cell can be resolved free from artefacts. However, ML-OS-SIM imaging was considerably faster: a 44~$\times$~44~$\times$~\unit[5]{$ \mu m^3 $} volume could be imaged in < \unit[2]{s}, whereas scanning confocal imaging took \unit[2]{min} \unit[36]{s}. That is, ML-OS-SIM was approximately 78 times faster. This speed advantage, in combination with the lower illumination laser power required, results in less photodamage and photobleaching to delicate samples (Supplementary note 1). An advantage of ML-OS-SIM is that a single detector can be used in conjunction with detection-splitting optics, whereas a confocal system would require an additional detector for each channel. 

\subsection{ML-OS-SIM enables the volumetric imaging of structures in multiple colours}

\begin{figure*} [htbp]
\centering
\includegraphics[width=0.99\textwidth]{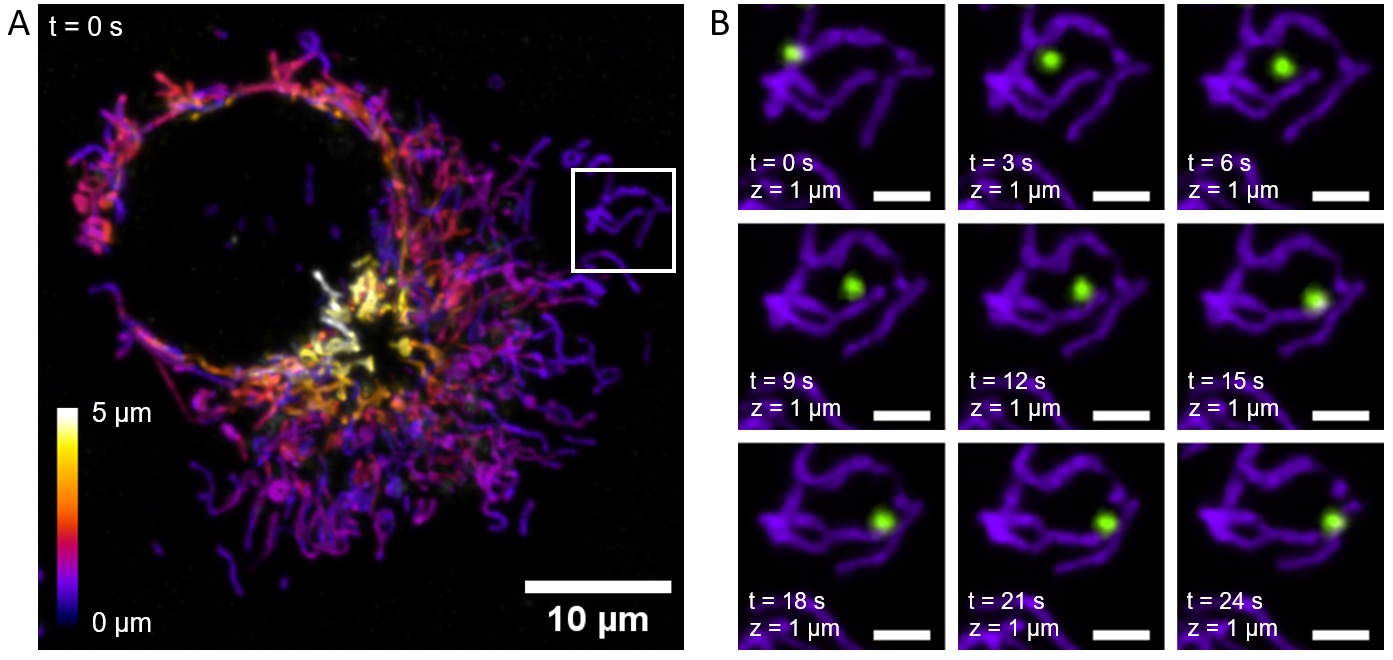}          
\caption{Machine learning optical sectioning structured illumination microscopy (ML-OS-SIM) enables video-rate mapping of the mitochondrial network in live cells. Live COS-7 cells were labelled with MitoTracker Orange and SiR Lysosome and excited simultaneously with the laser lines $\lambda=\unit[561]{nm}$ and $\lambda=\unit[647]{nm}$. A: Projection of volumetric data showing the mitochondrial network where the colourmap indicates the z position. A single channel is shown. Scale bar = \unit[10]{$\mu m$}. B: Dynamic interaction of lysosomes (green) with mitochondria (purple). Image sequence shows a projection of 3 adjacent slices from the data. Scale bars = \unit[2]{$\mu m$}.}
\label{fig:COS7_colourmap} 
\end{figure*}

To demonstrate the capabilities of the technique for biological imaging, live cells were imaged in multiple colours across a \unit[5]{$\mu m$} axial range (Fig. \ref{fig:COS7_colourmap}). Figure \ref{fig:COS7_colourmap}A depicts a maximum intensity projection of the mitochondrial network of a live cell imaged at a snapshot in time where the colourmap indicates the depth within the cell. No movement or striping artefacts are discernible in the image. Furthermore, the multicolour capability of the system allows several structures to be visualised without a temporal delay between colour channels. Figure \ref{fig:COS7_colourmap}B shows a two-channel time series of projections from the volumetric data. Here, a lysosome (green) can be seen to push a mitochondrial structure (purple) out of the way as it is transported through the cell. 

\subsection{Multicolour 3D imaging of DNA-nanotechnology biomimetic behaviours with ML-OS-SIM}

\begin{figure} [htbp]
\centering   
\includegraphics[width=0.99\textwidth]{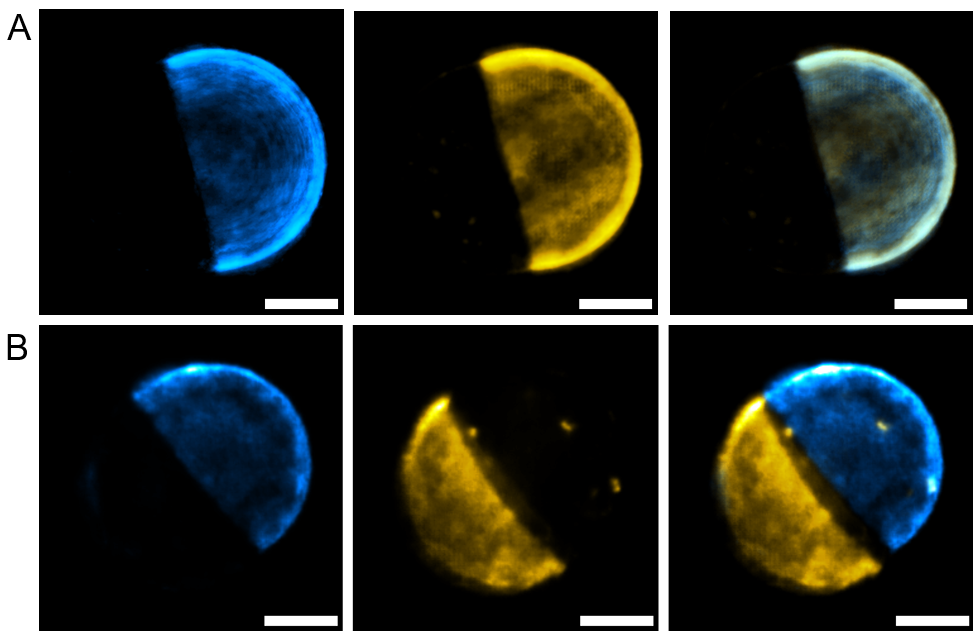}

\caption{ 
Optical sectioning and multicolour interferometric structured illumination enable fast, high-contrast 3D imaging of the lateral re-organisation of membrane-bound DNA nanostructures in Giant Unilamellar Vesicles (see Fig. S8) \cite{Rubio-Sanchez2021ADomains}. Images are 3D projections of volumetric data. A,B: Representative vesicles before (A) and 12~h after (B) fueling a biomimetic cargo transport pathway with nucleic-acid signals. A: In the initial configuration, the DNA (blue) and liquid disordered phase (yellow) overlap, indicating the preferential affinity of DNA nanostructures for liquid-disordered phases. B: Cargo transport relocates the fluorescent nanostructures to the liquid-ordered phase. ML-OS-SIM enableS samples to be imaged in multiple colours simultaneously, preventing motion artefacts and temporal offsets in the reconstructions. Scale bars = \unit[5]{$\mu m$}.} 

\label{fig:DNA bubbles} 
\end{figure}

In addition to biological samples, the contrast enhancement afforded by our ML-OS-SIM strategy enables the visualisation of processes in synthetic cell science. Recently, the synergy between lipid-membrane phase behaviour and the tools of DNA nanotechnology has gained traction to engineer bio-inspired responses in cell-like objects \cite{Langecker2014DNAMembranes, Rubio-Sanchez2021AmphiphilicBiology}. Figure \ref{fig:DNA bubbles} shows ML-OS-SIM images of DNA nanostructures tethered to phase-separated lipid membranes. The fast ML-OS-SIM technique allows for the reconstruction of multicolour 3D views of DNA-functionalised giant vesicles that can sustain biomimetic cargo transport pathways across their membrane surface (Fig. \ref{fig:DNA bubbles}) \cite{Rubio-Sanchez2021ADomains}. Chemical stimulus, in the form of DNA oligonucleotides, triggers the lateral re-organisation of DNA nanostructures, leading to the transport of fluorescently labelled DNA cargoes (blue), away from the liquid-disordered phase (yellow) to liquid-ordered lipid domains (Supplementary note 9). The responsive DNA nanostructures harness established strand displacement mechanisms \cite{Zhang2009ControlExchange, Zhang2011DynamicReactions} and the preferential tendency of different hydrophobic anchors in membrane-bound DNA to enrich distinct lipid domains \cite{Rubio-Sanchez2021ADomains}. The speed of ML-OS-SIM enables 3D imaging of the structures, allowing for better visualisation and understanding of the re-organisation process \cite{Rubio-Sanchez2021ADomains}. 

\section{Conclusion}

To date, OS-SIM has been limited by the unavailability of robust reconstruction algorithms and the inability to image live samples in multiple colours. Here, we demonstrate that these limitations can be overcome by combining ML reconstruction methods with interferometric SIM. We demonstrate fast, multicolour ML-OS-SIM at speeds of up to 37 Hz, across a 44~$\times$~\unit[44]{$\mu m^2$} field of view, with the option to view real-time reconstruction of OS-SIM data. The widefield detection method employed in ML-OS-SIM means that sectioning ability comparable to confocal microscopy can be achieved but at imaging speeds two orders of magnitude faster, with reduced photodamage to the sample. Additionally, we demonstrate that when combined with interferometric pattern generation, this can achieve multi-label imaging with no temporal lag, allowing for the interactions between structures to be visualised in 3D. These capabilities are typically only afforded by lightsheet imaging modalities but at the cost of greatly increased system complexity and non-conventional imaging geometries. OS-SIM with ML reconstruction offers a simpler alternative to lightsheet microscopy and is ideally suited to the imaging of dynamic samples in multiple colours. 

\section{Acknowledgments}
The authors would like to thank Lisa Hecker for her help constructing the interferometric setup. C.F.K. acknowledges funding from the UK Engineering and Physical Sciences Research Council (EP/L015889/1 and EP/H018301/1), the Wellcome Trust (3-3249/Z/16/Z and 089703/Z/09/Z), the UK Medical Research Council (MR/K015850/1 and MR/K02292X/1), and Infinitus Ltd. R.R.S. acknowledges funding from the Biotechnology and Biological Sciences Research Council through a BBSRC Discovery Fellowship (BB/X010228/1). R.R.S. and L.D.M. acknowledge funding from the European Research Council (ERC) under the Horizon 2020 Research and Innovation Programme (ERC-STG No 851667 NANOCELL). L.D.M. acknowledges funding from a Royal Society University Research Fellowship (UF160152, URF$\backslash$R$\backslash$221009). G.S.K.S. acknowledges funding from the Wellcome Trust (065807/Z/01/Z) (203249/Z/16/Z), the UK Medical Research Council (MRC) (MR/K02292X/1), Alzheimer Research UK (ARUK) (ARUK-PG013-14), Michael J Fox Foundation (16238 and 022159), and Infinitus China Ltd. S.K. acknowledges funding from the UK Engineering and Physical Sciences Research Council (EPSRC) grant EP/S023046/1 for the Centre for Doctoral Training in Sensor Technologies for a Healthy and Sustainable Future. R.M.M. and J.R.L. acknowledge funding from the UK Engineering and Physical Sciences Research Council (EP/S022139/1). 

\section{Code Availability}
Code available from https://github.com/edward-n-ward/ML-OS-SIM.

\bibliography{references}

\end{document}